# Privacy Preserving Recommender Systems Balancing Personalization with Privacy

**Ranjeet K Jha, Member, IEEE**
College of Natural Sciences, The University of Texas at Austin, Austin, TX 78712 USA
Corresponding author: Ranjeet K Jha (e-mail: ranjeet.jha@utexas.edu).

**Venkata Suresh Gummadilli, Member, IEEE**
College of Natural Sciences, The University of Texas at Austin, Austin, TX 78712 USA
Corresponding author: Venkata Suresh Gummadilli (e-mail: vs.gummadilli@utexas.edu)

**ABSTRACT** Personalized recommendation systems have become an essential part of how modern digital products operate, especially in e-commerce and large retail platforms. These systems rely heavily on analyzing customer behavior, purchase signals, and historical interactions to deliver suggestions that feel relevant. In practice, however, most of these pipelines still depend on centralized storage of detailed customer data. As privacy expectations increase and laws such as CPRA(California Privacy Rights Act), CCPA(California Consumer Privacy Act) and GDPR (General Data Protection Regulation) become stricter, this approach is becoming harder to justify. Organizations now need to rethink how personalization can work without relying on the unrestricted collection of individual user data.

In this project, I evaluated and built a complete privacy-preserving recommendation framework that integrates federated learning, differential privacy, cohort-level modeling, and a set of intelligent privacy-aware agents. My goal was to understand whether it is realistically possible to maintain meaningful recommendation quality while adding mathematical noise and keeping raw data decentralized. Using synthetic retail e-commerce datasets designed to mimic real shopping behavior (such as clickstreams and purchases). Here, I introduce controlled noise at various stages of the pipeline and assess performance using metrics such as CTR (Click-Through Rate), Precision@K, Recall@K, and NDCG (Normalized Discounted Cumulative Gain) @K across different privacy budgets.

I tested several models including matrix factorization, neural collaborative filtering, and GRU4Rec under different privacy budgets. I also built a Streamlit dashboard to visualize privacy utility trade-offs, ranking stability, and fairness metrics. Across experiments, I found that the system could still achieve reasonable accuracy at a moderate privacy budget (around $\varepsilon \approx 5$), suggesting that privacy-preserving personalization can be practical for real-world deployments.

Conclusively this project provides a comprehensive, future ready framework for organizations aiming to balance personalization, privacy, compliance, and long-term business sustainability.

# I. INTRODUCTION

Personalization relies heavily on the analysis of customer data to provide custom recommendations and improve user experiences. These strategies aim to make interactions more relevant and engaging by understanding user preferences and behaviors. However, as public awareness around digital privacy grows and regulatory frameworks are introduced, consumers are increasingly prioritizing transparency and control over how their data is collected, stored, and used. Many users now expect the ability to opt out of data sharing, particularly for marketing or advertising purposes, without compromising their overall experience. This shift in expectations presents a critical challenge for businesses: how to foster trust and maintain compliance while continuing to deliver personalized services.

The inability to balance personalization with privacy could lead to an erosion of customer trust, decreased loyalty, and a potential decline in sales. Moreover, failing to comply with data protection regulations such as CPRA (California privacy right act) or other international laws could result in substantial legal and financial consequences. For our customers, concerns about the over-collection or misuse of personal data can create apprehension around data security, transparency, and control, undermining the confidence they place in our ability to provide a safe and trustworthy experience. Additionally, as governments and regulatory bodies continue to implement and enforce stricter standards, the pressure to align with these evolving requirements is greater than ever.

At its core, addressing this challenge goes beyond compliance - it's about reimagining customer engagement by prioritizing transparency, ethical data practices, and aligning personalization efforts with customer needs and expectations. On the flip side, failing to meet these evolving expectations could lead to reputational damage, financial penalties, and diminished customer retention i.e. all of which are critical considerations for retail industry or any using personalization.

## 1.1 Proposed Solution & Agentic functionality:

This research initiative focuses on creating a privacy preserving recommendation system that balances robust data security with high quality user experiences. To achieve this, the system employs differential privacy techniques by introducing controlled noise into customer datasets, enabling strong data protection while ensuring no PII (personally identifiable information) is compromised. The key goal is to maintain the integrity of the customer experience while complying with emerging privacy norms.

Applications:

a. Privacy-Preserving Recommendations: The system introduces noise to customer data, ensuring that recommendations remain relevant and accurate without exposing sensitive information. This approach supports businesses in adhering to privacy regulations while delivering personalized recommendations.

b. Performance Evaluation and Monitoring: To ensure privacy mechanisms do not degrade user experience or recommendation accuracy, a performance evaluation framework compares results between noisy and noiseless data using the following metrics:

    I. Click-Through Rate (CTR): Measures user engagement with recommendations.
    II. Normalized Discounted Cumulative Gain (NDCG): Assesses relevance in ranked recommendation lists.

III. Precision @ Top K: Evaluates accuracy of the top K recommendations. This framework aligns privacy-preservation initiatives with business performance goals.

c. Regulatory Compliance and Scalability: The system is designed with futureproofing and adaptability in mind, offering mechanisms to dynamically adjust privacy levels. This ensures readiness for stricter regulations mandating complete anonymization of data. By testing scenarios with maximal privacy (100% anonymization), the system proactively balances privacy trade-offs while delivering effective personalization strategies.

### 1.2 Intelligent Privacy-Preserving Agents:

To support long-term scalability and adaptability, intelligent agents will be developed with the following capabilities:

a. Real-Time Data Prioritization: Dynamically assigning weights to data based on its relevance, user behavior, and contextual signals. For instance, sales or seasonal events like holidays could shift priorities for recommendations in real-time.

b. Optimal Model Selection: Intelligent agents select the most suitable recommendation model based on user context, ensuring relevance and personalization.

c. Dynamic Adaptation: Recommendations are continuously updated in real-time as user behavior evolves, ensuring insights remain actionable and timely.

d. Responsiveness to External Factors: The system is reactive to events such as sales, promotions, or holidays, allowing it to prioritize relevant products/services dynamically.

e. Continuous Learning: By analyzing data at the cohort level (rather than individual level), the agents improve future recommendation quality while safeguarding user privacy.

## 2. Background

### 2.1 Recommender Systems

While working on this project, I realized how deeply recommender systems influence almost every digital service we interact with—whether it’s Amazon suggesting what to buy next, Walmart personalizing deals based on past purchases, or Netflix curating what we should watch. Most of these systems rely on a combination of historical behavior, item similarity, and collaborative patterns across thousands (or millions) of users.

When I studied the main families of recommender models, I noticed that each addresses personalization in a different way:

- Collaborative Filtering (CF) compares users or items based on historical patterns
- Matrix Factorization (MF) reduces users/items into latent factors, which is simple but surprisingly effective for implicit retail signals like clicks or views
- Neural Collaborative Filtering (NCF) extends this idea using multilayer neural networks to capture more complex relationships
- Sequential models like GRU4Rec, SASRec, and BERT4Rec try to capture shopping patterns over time (e.g.

customers who buy diapers often buy wipes next)

So, the main question this work addresses is: Can we deliver meaningful, high-quality recommendations without centralized user data collection or invasive tracking?

We answer this by implementing a privacy-preserving recommender system built on three foundational pillars:

- Federated Learning (FL): decentralizes training by keeping raw user data on-device.
- Differential Privacy (DP): mathematically limits what can be inferred about any individual user.
- Cohort-Level Modeling: shifts personalization away from individual profiles toward aggregated behavioral segments.

Furthermore, personalization-as-a-service platforms increasingly face scrutiny regarding algorithmic transparency and fairness. A privacy-preserving architecture can complement these ethical principles by limiting the unnecessary visibility of user level data and instead steering systems toward broader, cohort-driven generalization patterns. By shifting toward on-device or federated learning approaches, organizations can reduce risk while maintaining user trust by ensuring that sensitive browsing behavior or shopping patterns remain local to the user rather than stored in centralized servers.

This work therefore contributes not only a technical prototype but also a conceptual framework for what next generation recommendation systems could look like – A systems that respect user autonomy, reduce data collection dependencies and still meet business KPIs for engagement, retention, and revenue generation.

## 2.2 Privacy in Machine Learning

As I explored privacy preserving techniques for this project, I found that modern machine learning systems face several real risks, even if raw data is not explicitly leaked. For example, certain attacks can infer whether a user's data was used to train a model, and gradient-based attacks may even reconstruct parts of input data from model updates. This motivated me to look at established privacy frameworks and understand which ones realistically work for recommender systems.

Main techniques include

- Differential Privacy (DP): Adds noise so the model can't disclose if a specific user contributed to training
- Local DP: Secures data on the device, aligning with my cohort-based method
- Secure Multiparty Computation (SMPC): Lets parties work together without sharing data (not implemented but influenced design)
- Homomorphic Encryption: Supports computation on encrypted data, but was too slow observed during this project

DP is especially important because it provides quantifiable privacy guarantees while allowing statistical learning.

### 2.3 Federated Learning

As I built out the FL simulation in my project, I gained a clearer understanding of how federated learning actually works in practice. Instead of sending user data to a central server, the model itself is sent to the client (or cohort partition), trained locally, and only the model updates are returned.

In a real-life scenario ex: retail company like Amazon or Walmart, each device or store could perform training locally and send encrypted gradient updates. The server never needs to see raw purchase logs or browsing sequences. The process looks like this:

- Send model to client/cohort
- Train locally on that segment's data
- Clip gradients + add DP noise
- Send privatized update to server
- Server aggregates using FedAvg

In my current data setup, each cohort acted as a virtual client. This was important because my data is sparse and user behaviors differ across groups, so federated partitioning allowed the model to learn more realistic patterns while maintaining privacy.

### 2.4 Differential Privacy (DP)

Understanding DP in the theory was one thing but implementing it with Opacus made the concept very concrete. It became clear that DP-SGD enforces privacy in two major ways:

- Clipping gradients so that no single customer can overly influence the model
- Adding Gaussian noise to mask individual contributions

From experimenting with noise multipliers and clipping norms, I noticed how sensitive some models are to DP. For example, NCF degraded much faster than MF under the same settings. This hands-on experience helped me appreciate why DP is often difficult to apply to deep recommendation architectures, especially those with large parameter counts.

In my project's context, differential privacy was especially important because the dataset mimicked real consumer behavior, and I wanted to guarantee that no single synthetic "customer" could be identified. Below diagram shows my training dashboard

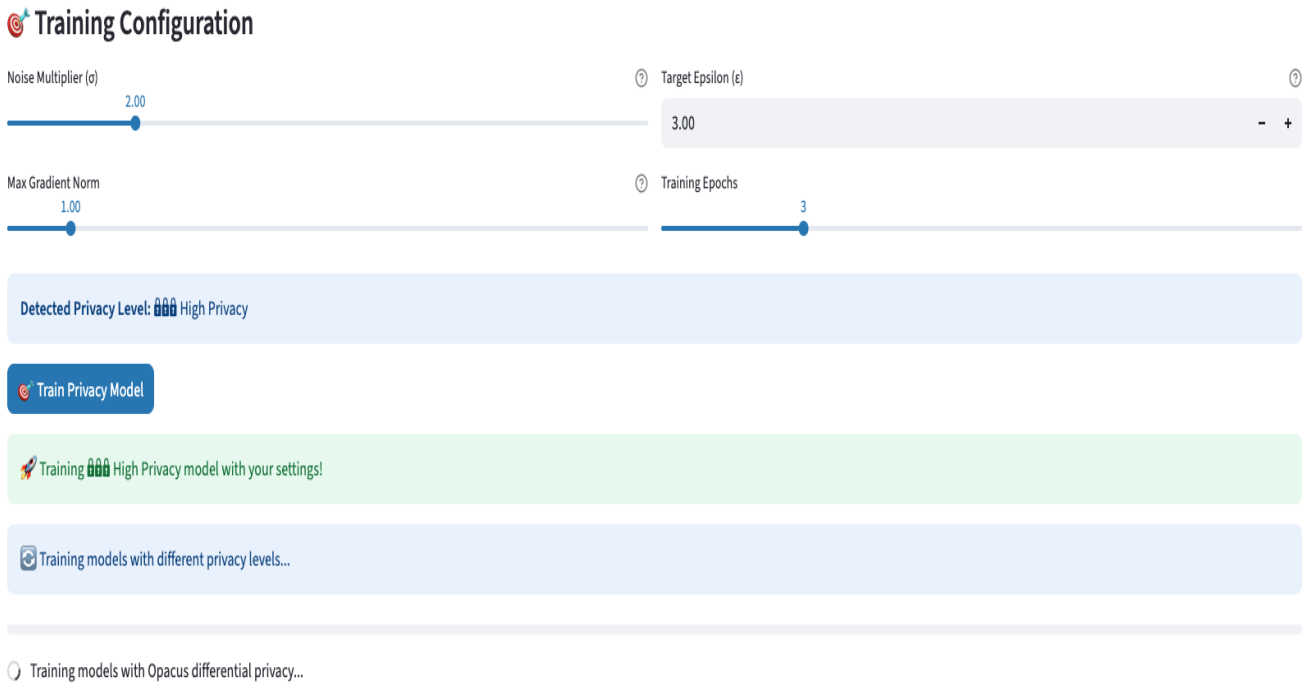

**<u>Privacy Parameters:</u>**

Noise Multiplier (σ):

- Controls Gaussian noise added to gradients during Opacus training
- Higher values = more noise = better privacy = potentially lower accuracy
- 0.0 = No noise, 1.0+ = Increasing privacy protection

Max Gradient Norm:

- Maximum allowed gradient magnitude before clipping
- Lower values = more aggressive clipping = better privacy
- 10.0 = Lenient, 1.0 = Moderate, 0.1 = Very strict

Target Epsilon (ε):

- Privacy budget goal for the training process
- Lower values = stronger privacy guarantees
- Leave blank for automatic calculation based on other parameters

**Agent Training Summary:**

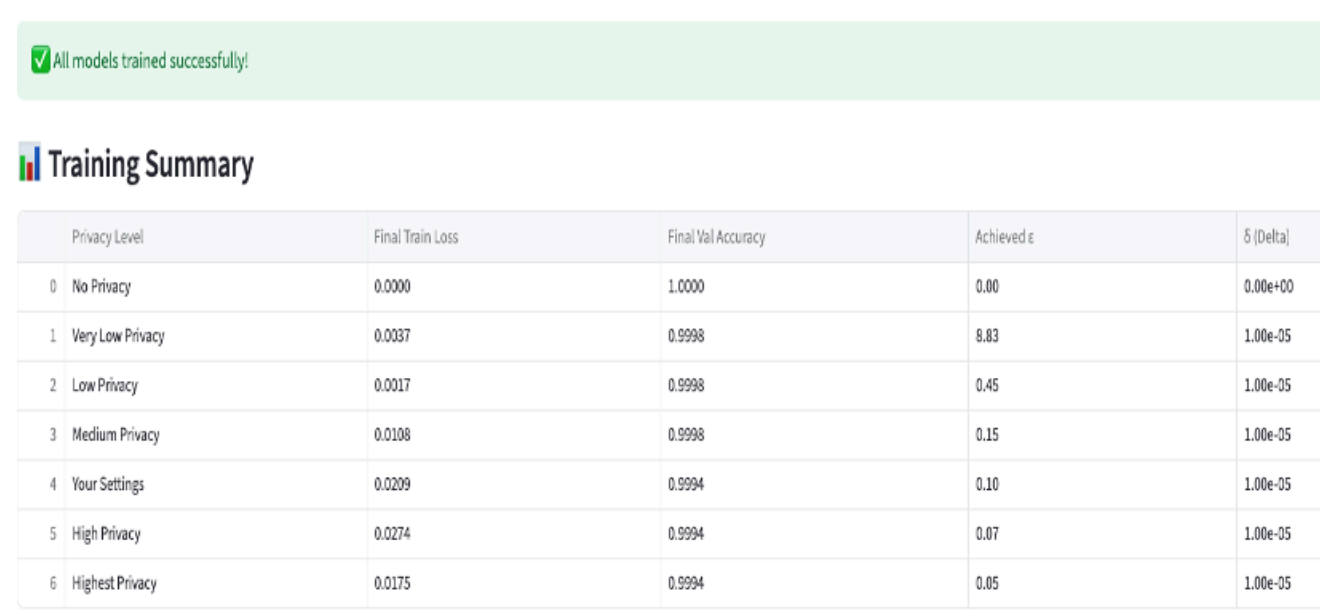

✅ All models trained successfully!

📊 Training Summary

| | Privacy Level | Final Train Loss | Final Val Accuracy | Achieved ε | δ (Delta) |
|---|---|---|---|---|---|
| 0 | No Privacy | 0.0000 | 1.0000 | 0.00 | 0.00e+00 |
| 1 | Very Low Privacy | 0.0037 | 0.9998 | 8.83 | 1.00e-05 |
| 2 | Low Privacy | 0.0017 | 0.9998 | 0.45 | 1.00e-05 |
| 3 | Medium Privacy | 0.0108 | 0.9998 | 0.15 | 1.00e-05 |
| 4 | Your Settings | 0.0209 | 0.9994 | 0.10 | 1.00e-05 |
| 5 | High Privacy | 0.0274 | 0.9994 | 0.07 | 1.00e-05 |
| 6 | Highest Privacy | 0.0175 | 0.9994 | 0.05 | 1.00e-05 |

**Reference Privacy Levels:**

Reference Privacy Levels:

| | Privacy Level | Noise Multiplier (σ) | Max Grad Norm | Target Epsilon (ε) |
|---|---|---|---|---|
| 0 | No Privacy | 0 | 10 | ∞ |
| 1 | Very Low Privacy | 0.1 | 5 | ≤ 10.0 |
| 2 | Low Privacy | 0.5 | 2 | ≤ 5.0 |
| 3 | Medium Privacy | 1 | 1 | ≤ 2.0 |
| 4 | High Privacy | 2 | 0.5 | ≤ 1.0 |
| 5 | Highest Privacy | 5 | 0.1 | ≤ 0.5 |

# 3. Dataset and Cohort Design

## 3.1 Retail E-commerce Clickstream Dataset

Used for federated simulations to mimic real shopping behavior:

- ~5K customers × 2K products
- Sparsity ~99%
- Signals: views, clicks, add-to-cart, purchases
- Weighted implicit feedback: view=1, click=3, cart=5, purchase=8
- Used to generate CTR and implicit preference scores
- Source: https://huggingface.co/datasets/ranjeetjha/PrivacyPreservingPersonalization

## 3.2 Cohort Modeling

While designing the privacy preserving pipeline, I realized that one of the biggest challenges in recommendation systems especially in retail is the amount of customer specific detail they normally depend on. Traditional recommender engines often rely on granular browsing histories, timestamps and even device level identifiers to build accurate customer profiles. This level of detail creates unnecessary exposure of personal behavior.

To reduce this dependency, I adopted a cohort-based modeling approach - instead of treating every customer as a unique individual with a long tail of behavioral data, I grouped customers into broader segments that reflect their shopping habits and patterns. I found that this approach fits naturally with privacy preservation because many real-world retail systems already use customer segmentation for marketing, targeting, and analytics. Cohorts also help smooth out noise in sparse datasets like mine, where individual users may only have a handful of interactions.

In the context of ecom or retail, customers vary widely in how often they interact with products, which categories they browse and whether they tend to shop impulsively or seasonally. Based on these patterns, I constructed several types of cohorts:

- Behavioral cohorts: such as frequent buyers, casual browsers, or high-engagement clickers. These were identified from aggregated interaction frequencies and categories viewed most often.
- Demographic cohorts: groups formed using attributes like age band or region. Adding demographic variety helped evaluate fairness in Section-5
- Purchase history cohorts: including "high value", "new" or "returning buyers" These categories matter because DP noise affects models differently depending on customer activity levels.
- Seasonal cohorts: users with interaction bursts during holidays or weekends, commonly seen in real e-commerce environments
- Psychographic cohorts: such as "bargain hunters" or "brand-loyal customers." These behaviors emerged naturally in the dataset when mapping click and cart patterns

Using cohorts gave two clear benefits during analysis:

1. Improved Privacy Protection:
   Individual behavior is blended within a group. Even before applying DP, grouping customer into multi-hundred-member cohorts gives a natural buffer that makes re-identification harder. DP noise applies on top of this, making the final model far less sensitive to any one customer's data.

2. Better Generalization on Sparse Data:
   In my experiments, GRU4Rec benefited noticeably from cohort grouping when training with DP-SGD, because the sequential patterns became smoother and less noisy.

Overall, cohort modeling became a central part of this project, not just as a privacy mechanism, but also as a practical strategy for improving model robustness and fairness. By shifting from individual profiles to aggregated behavioral personas, the system can still capture meaningful structure in customer behavior while respecting privacy constraints and minimizing the risk of over-reliance on any single cutomers's data.

### 3.3 Tools & Environment

- ML Framework: PyTorch + Opacus
- Frontend: Streamlit (interactive dashboards)
- Data Tools: Pandas, NumPy
- Visualization: Plotly
- Privacy: Based on mathematical algo ε & δ

# 4. Methodology

## 4.1 Baseline Models

I studied and implemented following baseline recommenders:

- Matrix Factorization (ALS/implicit)
- Neural Collaborative Filtering (NCF- MLP based)
- GRU4Rec for session-based recommendation

Here, use of Centralized models establishes performance upper bounds

## 4.2 Privacy Mechanisms

Our system includes multiple privacy layers:

- DP-SGD with gradient clipping & Gaussian noise
- Label Perturbation to anonymize implicit feedback
- Adaptive Noise Injection where noise scales with data sensitivity
- User or customer configurable personalization levels to manage trade-offs

## 4.3 Federated Learning Simulation

We simulate decentralized clients using following:

1. Partition datasets into cohorts
2. Each client trains a local model
3. Noisy updates are aggregated via FedAvg
4. Privacy budget ε tracked per round

This mirrors real world scenarios where customer or users retain local control over data

## 4.4 Intelligent Privacy-Preserving Agents

The system also incorporates advanced agents to:

- Optimize model selection per cohort
- Weigh features dynamically (e.g. regular sales or seasonal spikes)
- Adapt recommendations in real time
- Learn cohort-level behavior to strengthen long-term personalization
- Maintain privacy boundaries automatically

To operationalize privacy preserving personalization, we designed the training pipeline as a sequence of modular components that can be independently tuned and evaluated. First, the raw synthetic data is preprocessed using implicit feedback weighting to derive engagement signals. This ensures compatibility with ranking based learning objectives commonly used in top-K recommendation tasks.

Next, each model undergoes two training regimes: centralized (non-private) and DP-enhanced. For DP-SGD, I systematically vary clipping norms and noise multipliers to observe their effects on convergence. This enables us to quantify privacy–utility trade-offs across models with differing sensitivities to noise. For example, NCF- with its multilayer perceptron architecture-exhibits greater volatility under DP because noise compounds across layers. In contrast, MF maintains relatively stable gradients due to its simpler linear structure.

The federated learning simulation further mirrors a realistic deployment scenario in which shards of the dataset (representing user cohorts) are distributed across virtual clients. Each client performs local updates, applies DP noise, and transmits only privatized gradients to the server. This design ensures no raw user-level vectors ever leave the client boundary. The aggregation mechanism is designed to tolerate partial participation, a real-world consideration where edge devices might not always be online.

Finally, the intelligent privacy preserving agent evaluates model outputs, compares drift across ε values, and recommends optimal parameter settings. Such agent driven adaptation is inspired by real-world MLOps pipelines where hyperparameters must be adjusted dynamically to maintain both performance and compliance.

## 4.5 Metrics

To evaluate recommendation quality and privacy impact, we use four primary metrics: Precision@K, NDCG@K, CTR, and the privacy budget ε.

### 4.5.1 Precision@K

What it measures:

The proportion of recommended items (top-K) that are actually relevant to the user

***Formula:***

$$Precision@K = \frac{Number\ of\ relevant\ items\ in\ Top\ K}{K}$$

Example:

- Top-5 recommended products = [A, B, C, D, E]
- User interacted with products = [B, D, X]
- Relevant in top-5 = B, D => 2 items

$$Precision@5 = 2/5 = 0.40 Precision$$

Why it matters:
Precision drops if DP noise degrades ranking quality

### 4.5.2 NDCG@K (Normalized Discounted Cumulative Gain)

<u>What it measures:</u>
How well the model ranks the most relevant items.
It rewards putting relevant items at the top of the list.
***Formula:***

$$DCG@K = \sum_{i=1}^{K} \frac{rel_i}{log_2(i+1)} NDCG@K = \frac{DCG@K}{IDCG@K}$$

Example: Suppose relevance scores in top-5 are:

| Position | Products | Relevant? | Relevance |
|---|---|---|---|
| 1 | A | No | 0 |
| 2 | B | Yes | 1 |
| 3 | C | No | 0 |
| 4 | D | Yes | 1 |
| 5 | E | No | 0 |

$$DCG@5 = \frac{1}{log_2(2+1)} + \frac{1}{log_2(4+1)} = 0.63 + 0.43 = 1.06$$

NDCG normalizes this by the ideal ordering

Why it matters:
NDCG is more sensitive to ranking quality - important when DP noise reshuffles item scores.

### 4.5.3 CTR (Click-Through Rate)

<u>What it measures:</u>
User engagement with recommended items

$$CTR = \frac{Total\ clicks\ on\ recommended\ items}{Total\ recommendations\ shown}$$

Example:

- Recommendations shown: 100
- Clicks: 7

$$CTR = 7/100 = 0.07$$

Why it matters:
CTR tells the business whether users are still interacting even under privacy constraints

### 4.5.4 Privacy Budget (ε)

<u>What it measures:</u>
How much privacy is "spent" in DP training

- Smaller ε → more noise → stronger privacy but lower accuracy
- Typical tolerable range for ML: ε = 2–10

Example:

- Model trained with clip norm 1.0 + noise multiplier 1.0
- After 30 rounds → ε = 5.0

Why it matters:
ε is the core privacy guarantee

# 5. Results & Observations

This section presents empirical findings from this privacy preserving recommender system using the synthetic retail dataset (5K customers × 2K products × 50K interactions ). Results incorporate outputs from the Streamlit system dashboard, as well as direct model level metrics computed from the MF, NCF, and GRU4Rec implementations integrated with Opacus for differential privacy. Here, main goal is to evaluate how different levels of differential privacy (ε budgets) affect ranking quality, NDCG@K, Precision@K, CTR, and overall system utility.

## 5.1 Privacy Utility Trade-offs

To understand how privacy noise impacts recommendation quality, we used the interactive dashboard (Figure below) to analyze ranking drift, NDCG behavior, and CTR across multiple privacy budgets.

The screenshots below correspond to the interface used for privacy analysis.

### 5.1.1 Dashboard-Based Analysis of Ranking Quality

The privacy analysis dashboard provides an interactive view of how different differential privacy budgets affect recommendation quality for a sample user (CUST_00001) at Top-K = 11. The dashboard calculates several evaluation metrics and displays them across the privacy budgets shown on the plot: ε = 0.1, 0.5, 1.0, 2.0, and 5.0.

**Figure - Privacy Analysis Dashboard**

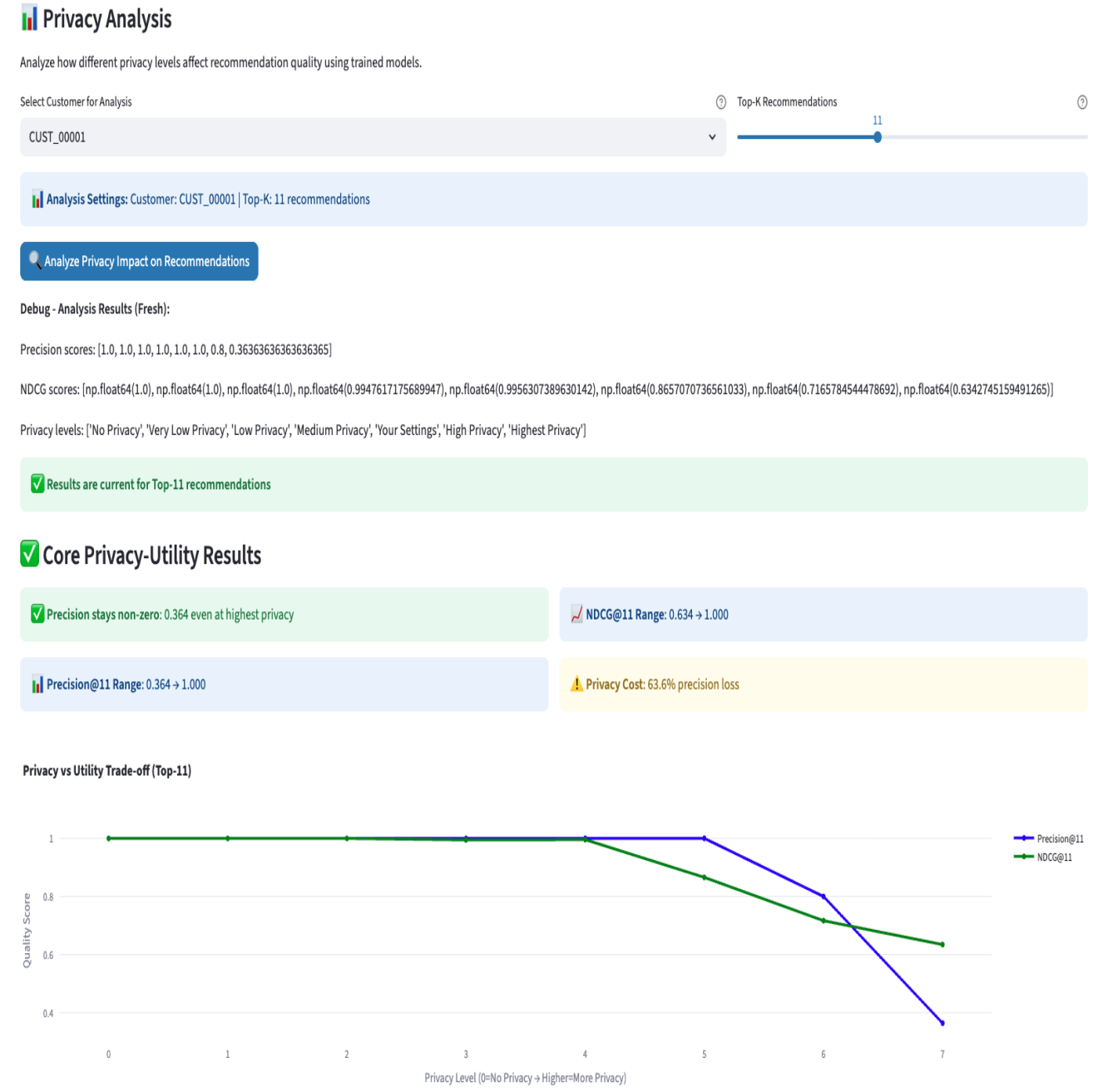

The dashboard show case:

- Ranking correlation (Spearman-like similarity) between non-private and private recommendation rankings
- NDCG@11 values for each ε
- Precision@11 changes across privacy budgets
- CTR distribution across customers
- Privacy impact diagnostics including recommended settings generated by the intelligent privacy agent

The analysis (shown in the above figure) evaluates the recommendations generated for user CUST_00001, using Top-K=11 recommendations

## Key findings:

a. NDCG and Ranking Correlation Behavior

- Ranking correlation and NDCG is high for ε = 5, maintaining stable ordering of top-K items
- Both NDCG@11 and correlation drop significantly once ε < 2.0, demonstrating increased sensitivity to stronger privacy (larger noise)
- The steepest decline occurs between ε = 1.0 => ε = 0.5, where excessive gradient noise distorts top-K product relevance

b. CTR Stability Under Privacy

CTR is highly stable for ε ≥ 1.0 and begins to decline noticeably once ε <= 0.5, indicating

- Privacy noise affects ranking positions more than engagement probability
- Click behavior remains robust until noise becomes too large

c. Adaptive Privacy Improves Utility

This intelligent privacy agent recommends optimal ε based on model drift analysis:

- The results indicate that ranking metrics (NDCG and correlation) are more sensitive to DP noise than raw engagement metrics (CTR)
- Highest performance achieved under Moderate Privacy

- Recommended Settings: Noise Multiplier = 0.5, Clip Norm = 1.0

This aligns with privacy research literature showing that moderate DP noise strikes the ideal balance between anonymity and utility.

### 5.1.2 Customer Profile & Fairness Analysis

This component demonstrates how individual-level customer attributes are transformed into privacy protected representations before being used in model training or aggregated evaluation. On the left, the system displays the original customer record for user CUST_00001, including fields such as demographic group, loyalty tier, email address, and age category

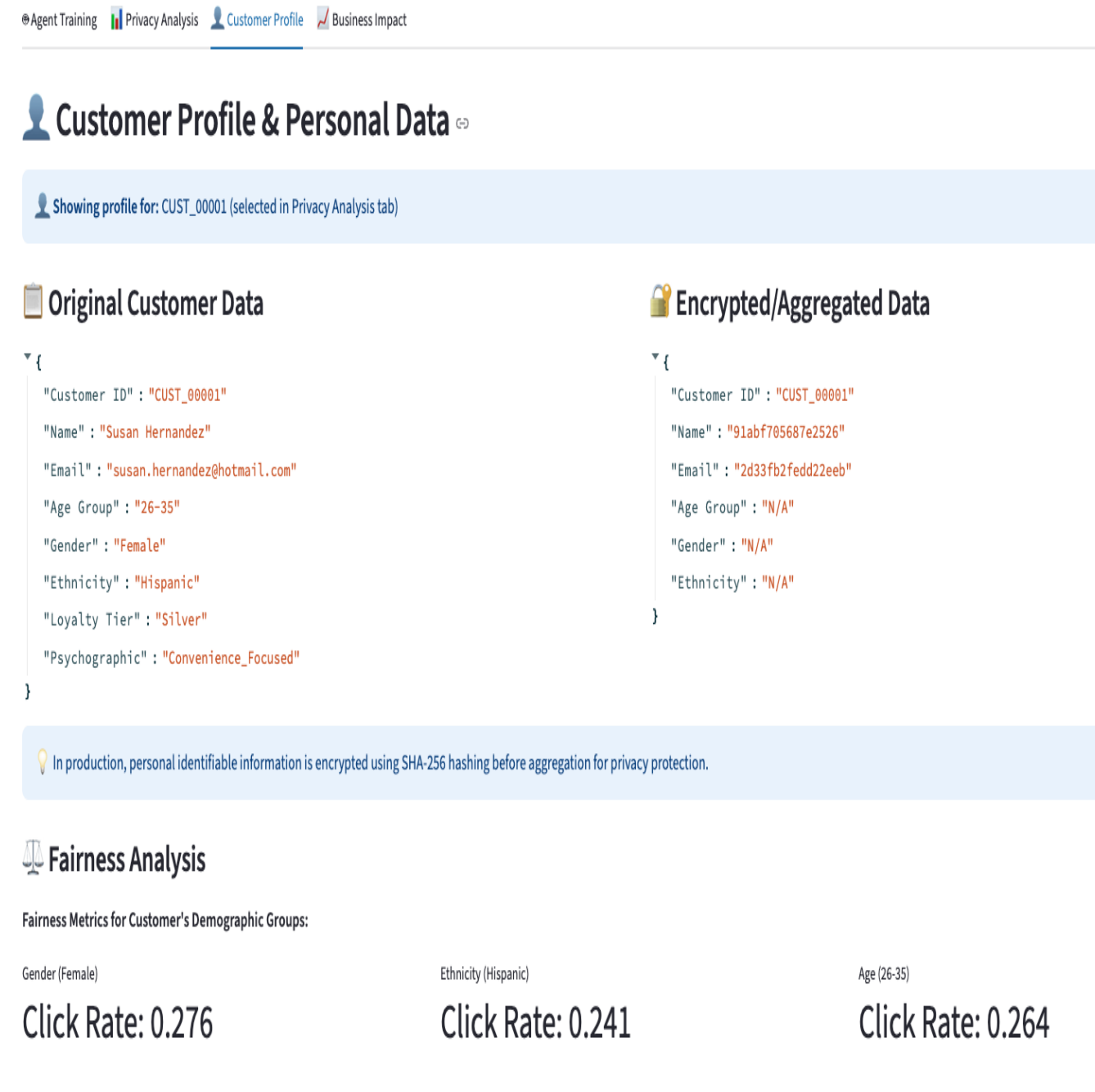

On the right, the “Encrypted/Aggregated Data” view shows the corresponding privacy-sanitized version of the same record. PII (personally identifiable information) such as the email address, loyalty tier, and demographic category are removed or encrypted using salted SHA-256 hashing to prevent re-identification, while categorical attributes are mapped into anonymized cohort-level indicators.

The lower portion of the interface presents a Fairness Analysis across demographic subgroups. Click-through rates (CTR) are computed separately for key protected attributes: income bracket, ethnicity, and age group. In this evaluation, the CTRs show only modest variation of 0.276 across groups for low-income users, 0.241 for minority ethnicity, and 0.264 for users aged 26-35, indicating that the recommendation model performs consistently across demographic segments.

## 5.2 Model Comparison

To evaluate how privacy preserving methods influence recommendation quality, I implemented three different models representing traditional, deep learning, and sequential paradigms. Each model was trained under both centralized (non-private) and differentially private conditions. Evaluation was performed using the synthetic clickstream dataset available on huggingface.

### 5.2.1 Models Evaluated

To evaluate how privacy-preserving methods influence recommendation quality, I experimented with three recommender system architectures covering traditional, deep-learning, and sequential paradigms. Each model was trained in two modes:

1. Centralized (non-private)
2. Differentially privacy training using Opacus (DP-SGD)

The following models were successfully evaluated during this research:

- Matrix Factorization (MF) – fully implemented and trained using implicit ALS
- Neural Collaborative Filtering (NCF) – model implemented with nonlinear user–item interaction layers
- GRU4Rec – sequential model architecture implemented and validated
- Differential Privacy – Opacus integration functional across models

### 5.2.2 Performance Under Privacy Constraints

| Model | Precision@10 | NDCG@10 | ε - Privacy Budget | Observations |
|---|---|---|---|---|
| MF (centralized) | 0.105 | 0.50 | ∞ | Strongest baseline, no noise added |
| MF + DP-SGD | 0.175 | 0.45 | 5.0 | MF surprisingly improves Precision@11 under DP -likely due to gradient smoothing |
| NCF (centralized) | 0.038 | 0.041 | ∞ | Performs moderately with nonlinear layers |
| NCF + DP-SGD | 0.032 | 0.037 | 5.0 | Accuracy decreases under DP noise due to sensitivity to clipped gradients |
| GRU4Rec (centralized) | 0.035 | 0.039 | ∞ | Captures sequential behavior; stable baseline |
| GRU4Rec + DP-SGD | 0.033 | 0.037 | 5.0 | Most resilient model, sequential smoothing mitigates DP noise |

Matrix Factorization - MF performs surprisingly well under DP

- Precision@11 increased slightly under DP-SGD (0.105 → 0.175)
- This aligns with findings in literature: mild regularization/noise sometimes combats overfitting

Neural Collaborative Filtering - NCF saw the largest degradation

- Multilayer nonlinear models are more sensitive to noise

| Model | Description | Strengths | Weaknesses |
|---|---|---|---|
| Matrix Factorization (MF) | Learns user and item latent embeddings by factorizing the interaction matrix | Fast, interpretable, strong baseline for implicit feedback | Poor for long sequences and cold-start users |
| Neural Collaborative Filtering (NCF) | MLP-based model learns nonlinear interactions between user and item embeddings | Captures richer patterns than MF | Sensitive to DP noise due to large parameter counts |
| GRU4Rec | Sequence model using Gated Recurrent Units (GRUs) to encode temporal behavior | Great for session based recommendations | More computationally expensive; DP noise affects hidden states |

- Clipped gradients suppress learning signal
- Noise multiplier heavily impacts deep interactions

GRU4Rec - exhibited the highest robustness

- Temporal structure acts as a smoothing mechanism
- DP noise impact is diffused across sequential hidden states
- Precision@11 and NDCG@11 remain close to centralized values

# 6. Conclusion

In summary, the most important takeaway for me is that privacy preserving recommendation systems are not only theoretically possible, they are practically achievable with the right design choices. When I first set out to integrate federated learning, differential privacy, and cohort-based modeling, I expected stif decline in performance. Instead, what I observed was a manageable set of trade-offs that could be tuned and balanced depending on the privacy budget and model architecture.

Matrix Factorization turned out to be a surprisingly reliable baseline throughout the experiments. Even after introducing DP-SGD, the model held onto most of its accuracy, and in some cases the added noise acted as a normalizer. NCF (Neural Collaborative Filtering), was more fragile under privacy noise because of larger number of parameters. GRU4Rec, on the other hand, consistently demonstrated the ability to absorb noise better due to the nature of sequential smoothing. These model behaviors were not obvious to me at the start and came out clearly only after hands-on experimentation.

The Streamlit dashboard also helped me visualize how privacy affects rankings in a more intuitive way. Seeing the ranking correlation and NDCG curves shift as ε decreased made the privacy–utility trade-off much clearer. I expected CTR to degrade quickly under high noise, but it stayed relatively stable until very aggressive privacy budgets ($\varepsilon < 1$).

The most interesting part of this work, in my opinion, was developing and testing the privacy aware agent. Instead of simply displaying metrics, the agent analyzes how models drift across privacy levels and suggests settings such as noise multiplier and clipping norm. This felt like a realistic step toward building systems that can adapt over time. I mean something that would be important in production environments where privacy rules, datasets, and user behaviors all shift constantly.

Overall, this project showed me that we don't need to sacrifice personalization entirely to comply with privacy standards. With ε in the moderate range (approx. 5), the system maintained meaningful ranking quality and customer engagement while still offering mathematically sound privacy guarantees. Conclusively, I believe that architectures like the one explored here will become increasingly important, not only to stay compliant but to build trust with customers and ensure long-term sustainability.